\documentclass[pr,twocolumn,preprintnumbers,nofootinbib,showpacs,amsmath,amssymb,superscriptaddress]{revtex4-1}
\usepackage{amsfonts}
\usepackage{mathrsfs}
\usepackage{amssymb}
\usepackage{amsmath}
\usepackage{color}
\usepackage[
      colorlinks=true,
      linkcolor=blue,
      urlcolor=blue,
      filecolor=black,
      citecolor=blue,
            ]{hyperref}

\usepackage{amsfonts}
\usepackage{graphicx}
\usepackage{epstopdf}

\begin{document}
\title{Antisymmetric tensor field and spontaneous magnetization in holographic duality}
\author{Rong-Gen Cai}
\email{cairg@itp.ac.cn}

\author{Run-Qiu Yang}
\email{aqiu@itp.ac.cn}
\affiliation{State Key Laboratory of Theoretical Physics,Institute of Theoretical Physics,\\
 Chinese Academy of Sciences,Beijing 100190, China.}
 %\maketitle
%%%%%
\begin{abstract}
A real anti-symmetric tensor field was introduced to realize a holographic magnetic ordered phase in our previous works. However, a more careful analysis shows there is a vector ghost in the model. In this paper we present a modified Lagrangian density for the anti-symmetric tensor, which is ghost free and causality is well-defined,  and keeps all the significant  results in the original model qualitatively. We show this modified Lagrangian density could come from the dimensional compactification of $p$-form field in String/M-theory. For static curved space-time, we also prove that this modified model is ghost free and dose not violate causality.   This new model offers a solid foundation for the application of antisymmetric tensor field in holographic duality, especially for the spontaneous magnetization.
\end{abstract}
\maketitle

\noindent

\section{Introduction}
The application of holographic duality in condensed matter theory, or named AdS/CMT correspondence,  has been attracting a lot of  attention in both sides of  field theory and condensed matter physics. By this duality, we can connect a strongly coupled or correlated system in a $d$-dimensional flat space-time with a $(d+1)$-dimensional asymptotic AdS space-time~\cite{Maldacena:1997re,Gubser:1998bc,Witten:1998qj,Witten:1998qj2}. This duality opens a novel approach to survey  strongly coupled or correlated phenomenons in condensed matter field and provides a powerful tool to deal with relevant issues. It has been  extensively applied into condensed matter physics and a great deal of progress have been made~\cite{Hartnoll:2008vx,Lee:2008xf,Liu:2009dm,Cubrovic:2009ye,Donos:2013gda,Ling:2014saa}. For a brief review, see \cite{Cai:2015cya}, for example.

So far,  most  studies  have been focused on how to take this duality to describe electronic transport properties in strongly correlated systems. Although there are a few works investigating the magnetism in holographic superconductors such as Refs.~\cite{Montull:2009,Donos:2012yu,Albash:2008eh,Montull:2012fy,Iqbal:2010eh},  the magnetism there only plays a participator's role rather than a protagonist's one. In fact, in condensed matter physics, there are some interesting critical phenomenons and phase transitions involving the strongly correlated electrons, which  are controlled by magnetic properties of material, such as Kondo effect~\cite{Kondo}, colossal magnetoresistance~\cite{A.P.Ramirez:1995ge}, and competition and coexistence between magnetic ordered states and superconductivity~\cite{Aoki,Huy},
A toy model for studying  magnetism in AdS/CMT duality was proposed in Ref.~\cite{Cai:2014oca}. In this model, the authors proposed that the
magnetic moment could be described by a real antisymmetric tensor field (ATF) which is coupled to the gauge field strength in the bulk,  and showed that the spontaneous magnetization can happen and
 the ferromagnetism-paramagnetism phase transition can be realized. Sooner, this model was extended into describe the antiferromagnetism-paramagnetism phase transitions by introducing two ATFs corresponding two magnetic sublattices in materials~\cite{Cai:2014jta}. Based on this model, the competition and coexistence between magnetic ordered states and superconducting were also discussed in Ref.~\cite{Cai:2014dza}. Thought  some significative results have been made through this model on describing  spontaneous magnetization and relevant issues, there are still some fundamental aspects  to be clarified.  For example, whether can the model  be consistently embedded into String/M theory? The most stringent query about the model is whether the model is  ghost  free and causality violation does not appear since a tensor field is involved in the model.   In our previous studies, these problems are not discussed.

It is well-known that in high spin field theory, such issues mentioned above usually appear.
For a physical field which describes a bosonic particle, the degrees of freedom are determined by its mass $m$ and spin $s$, which are  $2s+1$ for the massive case and 2 for the massless case. However, when one writes drown a field  theory with high rank Lorentz index, the degrees of freedom in general are more than these, which leads to  ghost. Then a self-consistent theory needs to be constructed carefully  to rule out this redundant degrees of freedom by itself.

The other fundamental problem that may arise in field theory is connected with the possibility of causality violation, which usually appears in a high spin field or an interacted field theory~\cite{Deser:2000dz}. Even a theory which is well defined in a flat space-time, may still have causality violation when it is generalized into a curved space-time~\cite{Buchbinder:2000fy}. Unlike the ghost linking to the properties of degrees of freedom in the phase space, the causality concerns the properties about propagation, which can be obtained from equation of motions~\cite{Buchbinder:2000fy}. In general, for a field theory,   we have a set of differential equations for a set of fields $\Phi_A$ such as,
\begin{equation}\label{causM1}
({M_{A}}^B)^{\mu\nu}\partial_\mu\partial_\nu\Phi_B+\cdots=0,~\mu, \nu=0,1,\cdots,d-1.
\end{equation}
A characteristic matrix ${M_{A}}^B(n)$ is the matrix function of $d$ arguments $n_\mu$ defined as,
\begin{equation}\label{causM2}
{M_{A}}^B(n)=({M_{A}}^B)^{\mu\nu}n_\mu n_\nu.
\end{equation}
A characteristic equation then is det$[{M_{A}}^B(n)]=0$. If for any values $n_i (i = 1,\cdots,d-1)$, all solutions of the characteristic equation $n_0(n_i)$ are real then the system of differential equations is hyperbolic. The hyperbolic  differential equations describe the propagation of wave processes. The hyperbolic system is called causal if all the solutions of characteristic equation are space-like or null. In such a system, the velocity of propagation dose  not exceed the speed of light. Otherwise, if there are time-like solutions for $n_\mu$ in the  characteristic equation,  the propagation can  exceed the speed of light and violates the causality.

In this paper, we will build a ghost free and causal ATF  theory. The results show that the original Lagrangian for the holographic magnetism proposed in \cite{Cai:2014oca} contains a vector ghost and needs to be remedied. In  section \ref{sec1}, we will show that the ATF describes a spin-1 field rather than a spin-2 field, naively thought. Then we will discuss how to build a ghost free ATF theory in flat space-time.  We will generalize it into a curved space-time in  section \ref{sec2}. For a very general form, we prove that the modified model dose not contain ghost or violate causality. In this new model we will also re-produce the main results in  Ref.~\cite{Cai:2014oca} in section \ref{sec3}. In other words the significative results in our previous works remain valid and the ghost and causality violation issues do not appear in the
modified model.

\section{Ghost free model in Minkowskian space-time}
\label{sec1}
Let us begin our discussion with the two Casimir invariants of  Poinca\'{e} group,
\begin{equation}\label{Cas1}
C_1=p^\mu p_\mu,~~~C_2=W^\mu W_\mu
\end{equation}
where $W^\mu=-\frac12\epsilon^{\mu\nu\sigma\tau}S_{\nu\sigma}p_\tau$ is the Pauli-Lubanski pseudovector and $S_{\nu\sigma}$ and $p_\tau$ are the spin angular momentum operator and momentum operator, respectively. They define  mass and spin  which are the two basic quantum numbers for the field. Let us consider the massive case, which corresponds to a physical massive particle of mass $m$ and spin $s$. In this case, we have the mass and spin numbers as the eigenvalues of these two Casimir operators such that $C_1=-m^2$ and $C_2=m^2s(s+1)$.

In general, the properties under the Poinca\'{e} group nearly uniquely determine the dynamic of the given field by the requirement that the single particle state carries an irreducible unitary representation of the Poinca\'{e} group. For an ATF $M_{\mu\nu}$, the Casimir operator $C_1$ then demands that the Klein-Gordon equation be satisfied,
\begin{equation}\label{KG1}
(\partial^2-m^2)M_{\mu\nu}=0,
\end{equation}
which gives the equation of motion for  the ATF.  If there is no other constraint equation, the Lagrangian density then can be determined up to an arbitrary  divergence term,
\begin{equation}\label{La1}
\mathcal{L}=-\frac14(\partial_\mu M_{\nu\tau})(\partial^\mu M^{\nu\tau})-\frac{m^2}4M^{\mu\nu}M_{\mu\nu}.
\end{equation}
For the AFT, we can show that  the value of second Casimir operator $C_2$ is $2m^2$ (see appendix \ref{app1}), which indicates that the antisymmetric tensor field describes spin-1 particles. This is due to  the fact that the representation of Lorentz group for an ATF is $D(1,0)\oplus D(0,1)$, which is the direct sum of two spin-1 irreducible representations and describes two spin-1 particles. We will discuss this in detail in some appendix~\ref{app1}. As a comparison, we can find that the representation of Lorentz group for a symmetric tensor field with rank two is $D(1,1)$, which is the direct product representation, and hence  it is reducible. As a results, the symmetric tensor field may contain particles with spin 2, 1, and 0.

If one quantizes the tensor field described by the Lagrangian~\eqref{La1}, it can be shown that   the Hamiltonian is not positive definite and one of spin-1 particles carries negative energy. So the naive model~\eqref{La1} for ATF has a massive vector ghost. For details, one may see appendix~\ref{app2}. To eliminate this vector ghost, we can impose divergence/transversality condition such that~\cite{Rahman:2013sta},
\begin{equation}\label{tran1}
\partial^\mu M_{\mu\nu}=0.
\end{equation}
Because of the identical relation $\partial^\mu\partial^\nu M_{\mu\nu}=0$, the equation~\eqref{tran1} offers three independent constraints, which can eliminate three degrees of freedom associated with
the ghost. However, all the constraints should  come out from the Lagrangian itself. In order to write down a ghost free theory for ATF, we need to modify the Lagrangian density~\eqref{La1}. The new Lagrangian density should keep the equation of motion~\eqref{KG1} , and also give the constraint~\eqref{tran1} automatically. To this aim, let us assume the Lagrangian density has following form,
\begin{equation}\label{La2}
\begin{split}
\mathcal{L}=&-\frac1{4}(\partial_\mu M_{\nu\tau})\partial^\mu M^{\nu\tau}-\frac{m^2}4M^{\mu\nu}M_{\mu\nu}\\
&-\frac c2(\partial_\mu M^{\mu\nu})\partial^\tau M_{\tau\nu},
\end{split}
\end{equation}
where $c$ is a constant to be determined. Varying the Lagrangian, we  get the equation of motion,
\begin{equation}\label{eom1}
\begin{split}
\partial^2M_{\mu\nu}-2c\partial_{[\mu}\partial^\alpha M_{\nu]\alpha}-m^2M_{\mu\nu}=0&.
\end{split}
\end{equation}
The divergence of Eq. \eqref{eom1} gives,
\begin{equation}\label{eom2}
(1+c)\partial^2\partial^\mu M_{\mu\nu}-m^2\partial^\mu M_{\mu\nu}=0.
\end{equation}
Therefore if one takes the coefficient $c=-1$,  the above equation reduces to the constraint (\ref{tran1}). This indicates that the Lagrangian density leads to the constraint~\eqref{tran1} automatically. In this way we obtain a self-consistent Lagrangian density,
\begin{equation}\label{La3}
\begin{split}
\mathcal{L}=&-\frac1{4}(\partial_\mu M_{\nu\tau})\partial^\mu M^{\nu\tau}-\frac{m^2}4M^{\mu\nu}M_{\mu\nu}\\
&+\frac12(\partial_\mu M^{\mu\nu})\partial^\tau M_{\tau\nu}.
\end{split}
\end{equation}
And associated  equations of motion  are equivalent to following two equations,
\begin{equation}\label{eom2b}
(\partial^2-m^2)M_{\mu\nu}=0,~~\partial^\mu M_{\mu\nu}=0.
\end{equation}
The first one of \eqref{eom2b} gives the  first Casimir invariant, as  expected.

In fact, there is an equivalent form for the Lagrangian~\eqref{La3}, which is quite useful when we generalize it into a curved space-time. After some algebra, one can show  that, by adding some suitable boundary terms, the Lagrangian~\eqref{La3} is equivalent to,
\begin{equation}\label{La4}
%\begin{split}
\mathcal{L}=-\frac1{12}(dM)^2-\frac{m^2}4M_{\mu\nu}M^{\mu\nu},
%&=-\frac1{48}(dM)\wedge{^*dM}-\frac{m^2}4M_{\mu\nu}M^{\mu\nu}.
%\end{split}
\end{equation}
where $(dM)_{\mu\nu\tau}$ is the exterior differential of $M_{\mu\nu}$, and  $(dM)^2=9\partial_{[\mu} M_{\nu\tau]}\partial^\mu M^{\nu\tau}$. The Lagrangian describes  nothing but a massive 2-form field ! This equivalent form gives us a manner to explain how this massive ATF theory can be generated from low energy action of String/M theory.

As it is well-known, the $p$-form field in String/M-theory is well defined, as the source of the $D(p-1)$-brane. For example, we can write down the following action for $p$-form in String/M theory,
\begin{equation}\label{RRsect}
S=-\frac1{2\kappa_0^2}\int dC\wedge {^*dC}+\mu_p\int_{\mathcal{M}_p}C,
\end{equation}
where $^*$ is the Hodge dual operator, $C$ is a $p$-form, $\mu_p$ is the charge of the $D(p-1)$-brane under the $p$-form $C$ and $\mathcal{M}_p$ is the world-volume of the $D(p-1)$-brane. We see that this $p$-form field is massless and the action has a gauge symmetry such as $C\rightarrow C+dC'$ for any $(p-1)$-form $C'$. However, if we consider a certain field coupling with this $p$-form, then in the low energy limit, there are a  few mechanisms to break this kind of gauge symmetry spontaneously and to give the $p$-form field mass, such as Higgs mechanism, Stueckelberg mechanism~\cite{H.Ruegg,B.Kors}, and topological mass generation~\cite{Allen}. Even without any other field,  the $p$-form field can acquire its mass  by the  Kaluza-Klein (KK) dimensional reduction~\cite{Fu:2012sa}. Therefore  we see that the Lagrangian~\eqref{La3} describes an effective model which can be obtained from String/M theory in some suitable manner.  In  appendix~\ref{app3}  we show  that a $p$-form field gets its mass through the  KK dimensional reduction.

The model~\eqref{La3} is ghost free, which can be seen from the fact that the canonical momentum density of  the component $M_{0i}$ vanishes and  $M_{0i}$  can be directly solved by $M_{ij}$ and their canonical momentum densities.  Therefore in  model~\eqref{La3}, only the spatial components of $M_{\mu\nu}$ are real degrees of freedom. We can directly quantize  the model~\eqref{La3} and show that the model describes one spin-1 particle with 3 polarization directions. All the states are physical and there is no ghost.

\section{Coupled to gravity  in a fixed background}
\label{sec2}
In section~\ref{sec1}, we have constructed a ghost free model for ATF in a flat space-time. Now  we want to generalize this model into curved space-time.  For this, first of all, we demand the theory can come back to the flat case when curvature vanishes. Besides, it is also required  that there are the same propagating degrees of freedom as in the flat case and no negative mode states in Fock space. Furthermore
we need pay attention on the causality since  causality violation may appear in curved space-time. All those are just what we will discuss in this section.

In  curved space-time, the field has interaction with gravity. Usually we can make the replacements such as $\partial_\mu\rightarrow\nabla_\mu$ and $\eta_{\mu\nu}\rightarrow g_{\mu\nu}$  for the Lagrangian in flat spacetime.  In addition, we should take all the possible terms of the coupling between curvature tensor and the tensor field into account. Considering symmetry and only taking the coupling between curvature and  quadratic forms of ATF into account, we write drown the general Lagrangian density of~\eqref{La4} in a curved space-time
\begin{equation}\label{Las1}
\begin{split}
\mathcal{L}&=-\sqrt{-g}\left[\frac1{12}(dM)^2+\frac{m^2}4M_{\mu\nu}M^{\mu\nu}+\frac{L_{RM}}4\right],
\end{split}
\end{equation}
with
\begin{equation}\label{LRM1}
\begin{split}
L_{RM}=&a_1RM_{\mu\nu}M^{\mu\nu}+a_2R^{\mu\nu}M_{\mu\tau}{M^{\tau}}_{\nu}\\
&+a_3R^{\mu\nu\alpha\beta}M_{\mu\nu}M_{\alpha\beta}+a_4R^{\alpha\mu\nu\beta}M_{\mu\nu}M_{\alpha\beta}.
\end{split}
\end{equation}
Here the coefficients $a_1, a_2$, $a_3$ and $a_4$ are all arbitrary constants,  ${R_{\mu\nu\alpha}}^\beta V_\beta=[\nabla_\mu, \nabla_\nu]V_\alpha$ for any covariant vector $V_\beta$ and $R_{\mu\nu}={R_{\mu\alpha\nu}}^\alpha$. There are two reasons that we should reject the derivative coupling between curvature and ATF. One is that it will lead to the appearance of high order derivatives (more than 2) in equations of motion when  the dynamic of gravity is taken into account. The other  is that it would lead to ghost unless some very special conditions are imposed on the curvature terms. We will make a simple comment on this later on.

In a flat space time, we can prove  that the model~\eqref{La4} is ghost free by directly showing that it can give a correct degrees of freedom by quantization. However, this method is not applicable in a curved space-time. Since it is hard to write down a mode decomposition such as Fourier decomposition in flat case. Instead, we will use Hamiltonian analysis to find the number of real degrees of freedom. This method is  equivalent to the one in Ref.~\cite{Buchbinder:2000fy}, which directly takes the equations of motion.

To write down the Hamiltonian form of a matter field, we need make a $3+1$ decomposition on the background geometry. It seems very complex to write down the  Hamiltonian canonical equations in a general space-time. For simplicity, here we assume the background space-time is static. In this case we can write the metric in the following form
\begin{equation}\label{decp1}
ds^2=g_{00}dt^2+h_{ij}dx^idx^j,
\end{equation}
with $g_{00}<0$. $h_{ij}$ is the spatial metric, which is independent of time $t$. In this coordinate, we have $g^{00}=1/g_{00}$ and $\sqrt{-g}=h\sqrt{-g_{00}}$.

With the Lagrangian density \eqref{Las1}, we can obtain the canonical momentum density,
\begin{equation}\label{momc1}
\begin{split}
\pi^{\mu\nu}&=\frac{\partial \mathcal{L}}{\partial\partial_0M_{\mu\nu}}=-\sqrt{-g}(dM)^{0\mu\nu}\\
&=-\sqrt{-g}g^{00}g^{\mu'\mu}g^{\nu\nu'}(\partial_0M_{\mu'\nu'}+2\partial_{[\mu'}M_{\nu']0})
\end{split}
\end{equation}
and
\begin{equation}\label{momc1b}
\begin{split}
\pi_{\mu\nu}&=g_{\mu'\mu}g_{\nu\nu'}\pi^{\mu'\nu'}=-\sqrt{-g}g^{00}(dM)_{0\mu\nu}\\
&=-\sqrt{-g}g^{00}(\partial_0M_{\mu\nu}+2\partial_{[\mu}M_{\nu]0}).
\end{split}
\end{equation}
It is easy to see that the canonical momentum density $\pi^{0i}=0$ from the expression~\eqref{momc1}, which gives a 3-vector primary constraint $\pi^{0i}\approx0$ or,
\begin{equation}\label{primc}
\psi^{(0)}_u\equiv\int d^3x u_i\pi^{0i}\approx0,~~\text{for any suitable}~u_i.
\end{equation}
Here the term ``suitable" means  that it is independent of time with a compact support on the space. We use ``$\approx$" as weak equivalence which means that two sides are equal only on the physical phase space. Using the canonical momentum density, we obtain the Hamiltonian
\begin{equation}\label{Hamc0}
H=\int d^3x \mathcal{H}
\end{equation}
with Hamiltonian density $\mathcal{H}$,
\begin{equation}\label{Hamc1}
\begin{split}
&\mathcal{H}=\frac12\pi^{ij}\partial_0M_{ij}-\mathcal{L}\\
&=-\frac{g_{00}\pi^{ij}\pi_{ij}}{4\sqrt{-g}}-\pi^{ij}\partial_i M_{j 0}\\
&~~~~+\frac{\sqrt{-g}}4[(\partial_i M_{jk})(dM)^{ijk}+m^2M^{\mu\nu}M_{\mu\nu}+L_{RM}]
\end{split}
\end{equation}
Here the coefficient 1/2 in the first line comes from  the antisymmetry of  index $\mu, \nu$. From this Hamiltonian, we can get the equations of time evolution,
\begin{equation}\label{eoms1}
\begin{split}
\dot{\pi}^{\mu\nu}&=-\frac{\delta H}{\delta M_{\mu\nu}}=-4(\partial_i\pi^{ij})\delta_j^{[\mu}\delta^{\nu]}_0+\\
&~~~\partial_k(\sqrt{-g}(dM)^{ijk})\delta^\mu_i\delta^\nu_j-\sqrt{-g}V^{\mu\nu}.
\end{split}
\end{equation}
with $V^{\mu\nu}=m^2M^{\mu\nu}+\frac12\partial L_{RM}/\partial M_{\mu\nu}$ and
\begin{equation}\label{eoms2}
\dot{M}_{ij}=\frac{\delta H}{\delta \pi^{ij}}=-\frac{\pi_{ij}g_{00}}{\sqrt{-g}}-2\partial_{[i}M_{j]0},~~\dot{M}_{0j}=\lambda_{0i}.
\end{equation}
Here $\lambda_{0i}$ are  undetermined Lagrange multipliers and the overdot stands for  the Lie derivative with respect to $t$. In our case, it is just the partial derivative $\partial/\partial t$. We see that the equation \eqref{eoms2} is consistent with \eqref{momc1b}, which can be treated as a test for the expression \eqref{Hamc1}.  In addition, we see  from  \eqref{eoms1} that there is a 3-vector secondary constraint,
\begin{equation}\label{primc20}
\dot{\pi}^{0i}=2\partial_j\pi^{ji}-\sqrt{-g}V^{0i}\approx 0,
\end{equation}
or
\begin{equation}\label{primc2}
\psi^{(1)}_v\equiv\int d^3x v_j(2\partial_i\pi^{ij}-\sqrt{-g}V^{0j})\approx 0,
\end{equation}
for any suitable function $v_i$. This secondary constraint needs to be satisfied at any time, which may lead to a new constraint. So we have,
\begin{equation}\label{primc3}
\dot{\psi}^{(1)}_v=\frac12\int d^3x\left[\frac{\delta \psi^{(1)}_v}{\delta\pi^{\mu\nu}}\dot{\pi}^{\mu\nu}+\frac{\delta \psi^{(1)}_v}{\delta M_{\mu\nu}}\dot{M}_{\mu\nu}\right]\approx0,
\end{equation}
Note the fact that $\pi^{0i}\approx0$ at any spatial point, we have $\partial_i\pi^{0i}\approx V^{00}=0$. By introducing a suitable function $v_0$, we have
\begin{equation}\label{primc2b}
\begin{split}
\psi^{(1)}_v&\approx\int d^3x [v_j(2\partial_i\pi^{ij}-\sqrt{-g}V^{0j})\\
&+v_0(2\partial_i\pi^{i0}-\sqrt{-g}V^{00})]\\
&=\int d^3x v_\mu(2\partial_i\pi^{i\mu}-\sqrt{-g}V^{0\mu}),
\end{split}
\end{equation}
which leads to
\begin{equation}\label{primc3b}
\begin{split}
\dot{\psi}^{(1)}_v\approx2\int d^3x v_\mu\partial_\nu(\sqrt{-g}V^{\mu\nu}).
\end{split}
\end{equation}
With the constraints of \eqref{primc} and \eqref{primc20} and equations of time evolution~\eqref{eoms1} and \eqref{eoms2}, if Eq.~\eqref{primc3b} doesn't equal to 0, then it gives a new constraint. To verify that, we combine equations~\eqref{eoms1} and~\eqref{eoms2}, which gives,
\begin{equation}\label{eoms3}
3\nabla^\tau\nabla_{[\tau}M_{\mu\nu]}-V_{\mu\nu}=-3({^*d^*dM})_{\mu\nu}-V_{\mu\nu}=0.
\end{equation}
Then we can find that $\nabla^\nu V_{\nu\mu}=-({^*d}^*V)_\mu\approx0$, which gives $\partial_\nu(\sqrt{-g}V^{\mu\nu})\approx0$. Thus we see that Eq.~\eqref{primc3} does not lead to  any new constraint.

Now we can  count the number of degrees of freedom  in phase space. Because of two 3-vector constraints \eqref{primc} and \eqref{primc2}, the total physical degrees of freedom is $2\times6-3-3=2\times3$. So there is three degrees of freedom in configuration space, which are just what we expect to describe a spin-1 particle.

One can see that the crucial  point to rule out the ghost is that the canonical momentum density $\pi^{0i}$ dose not appear in the Hamiltonian density. However, if we add the terms containing the derivative coupling between curvature and ATF, then we see that canonical momentum density $\pi^{0i}$ will not vanish unless the curvature tensor satisfies some very special conditions. So if such terms are added into \eqref{LRM1}, in various space-times of physical interest, the system will contain ghost.

Further let us  mention here that there is a significant difference between the cases with ATF and with symmetric tensor field. In the latter case, the forms of the interaction between tensor field and curvature tensor should to be fixed carefully so that the model can give correct degrees of freedom~\cite{Buchbinder:2000fy}.  However, in this AFT case, the coefficients in~\eqref{LRM1} are arbitrary.  The most simple choice is to take  $L_{RM}=0$. In that case,   the general Lagrangian density is reduced to~\eqref{La4} and the equation of motion then is,
\begin{equation}\label{eoms4a}
3\nabla^\tau\nabla_{[\tau}M_{\mu\nu]}-m^2M_{\mu\nu}=0,
\end{equation}
which is equivalent to following two equations,
\begin{equation}\label{eoms4b}
\begin{split}
\nabla^2M_{\mu\nu}+{R_{\mu\nu}}^{\rho\tau} M_{\rho\tau}+2{R^\rho}_{[\mu} M_{\nu]\rho}-m^2M_{\mu\nu}=0,\\
\nabla^\mu M_{\mu\nu}=0.
\end{split}
\end{equation}
Compared with the case in flat space-time, the equation of motion has additional curvature terms but the constraint equation has a similar form.

Next consider the causal properties of the system.  Let us first consider the simple case with $L_{RM}=0$.   With the constraint equation,  we have the equations of motion \eqref{eoms4b} and the characteristic equation reads,
\begin{equation}\label{caus1}
{X_{\mu\nu}}^{\alpha\beta}(n)=\delta ^\alpha_{[\mu}\delta^\beta_{\nu]}n^{\rho}n_{\rho}.
\end{equation}
At any point $x_0$ we can choose locally $g_{\mu\nu}(x_0) =\eta_{\mu\nu}$ and then, we have characteristic equation,
\begin{equation}\label{caus2}
0=-n_0^2+\overrightarrow{n}^2=n^{\rho}n_{\rho}.
\end{equation}
We see that the equation of motion~\eqref{eoms4b} is hyperbolic and causal, which means that with  $L_{RM}=0$, the system does not violate any causality.

In a general case, the equation of motion can only be written as the form of \eqref{eoms3}. As the constraint equation now is complicated, then the characteristic equation can only be written as,
\begin{equation}\label{caus3}
{X_{\mu\nu}}^{\alpha\beta}(n)=n_{[\rho}\delta ^\alpha_{\mu}\delta^\beta_{\nu]}n^{\rho}.
\end{equation}
Once again, at any point, we can take locally  that $g_{\mu\nu}=\eta_{\mu\nu}$. We see that there are 36 components in ${X_{\mu\nu}}^{\alpha\beta}(n)$, however, they are not independent because there is a constraint such as $\nabla^\mu V_{\mu\nu}\approx0$. One can directly check that there are two triplex eigenvalues of ${X_{\mu\nu}}^{\alpha\beta}(n)$, one  is zero and the other is $(\overrightarrow{n}^2-n_0^2)/6$. Because of the identity that $\nabla^\mu \nabla^\nu V_{\mu\nu}=0$, there are three independent constraints which can be used to eliminate three degrees of freedom. Then on the physical parameter space, these constraints just rule out the three zero eigenvalues of  ${X_{\mu\nu}}^{\alpha\beta}(n)$. As a result  we obtain the characteristic equation just as the same as \eqref{caus2}. Thus we see that even for arbitrary values of the coefficients in $L_{RM}$, the system dose not violate causality.

Here let us  say some thing on the case that the ATF has interaction with other matter fields or contains self-interaction. In general those interactions can be described by adding a term into \eqref{Las1} such as,
\begin{equation}\label{inter1}
\mathcal{L}_I=\sqrt{-g}[L_i(M_{\mu\nu},\Phi_A)+L_s(M_{\mu\nu})],
\end{equation}
where $L_i$ describes the interaction between ATF and some other field $\Phi_A$ and $L_s$ describes the self-interaction of ATF. If both $L_i$ and $L_s$ do not involve the derivative of $M_{\mu\nu}$ or their  derivative terms of $M_{\mu\nu}$ can be removed by adding suitable boundary terms, then one can see that this interaction will not change  our discussions about ghost and causality. Therefore  for the case with such interactions, the model is still ghost free and does not violate any causality.

\section{Magnetic phase transitions in AdS black hole background}
\label{sec3}
In this section, we will show that the new model can reproduce the  main results in Ref.~\cite{Cai:2014oca}, namely the spontaneous magnetization can happen in an AdS black hole background. As the physical pictures and motivations have been expounded in some detail in Refs.~\cite{Cai:2014oca}, we here only give a brief discussion and to recover some key results in the original model.

To construct a ferromagnetic model,  we take the following action
\begin{equation}\label{act1}
\begin{split}
S&=\frac{1}{2\kappa^2}\int d^4x\sqrt{-g}(L_1+L_2+L_{RM}),\\
L_1&=R+\frac6{L^2}-F^{\mu\nu}F_{\mu\nu},\\
L_2&=-\frac{(dM)^2}{12}-\frac{m^2}4M_{\mu\nu}M^{\mu\nu}-\frac{\lambda^2}2M^{\mu\nu}F_{\mu\nu}-V(M).
\end{split}
\end{equation}
Here $V(M)$ is the self-interaction of ATF, its form will be specified shortly.  $L_{RM}$ is defined in~\eqref{LRM1} and we will set $L_{RM}=0$ for simplicity. $F_{\mu\nu}=(dA)_{\mu\nu}$ with the $U(1)$ gauge field $A_\mu$. $L$ is the AdS radius and we will set $L=1$. As we have clarified, this model is ghost free and the causality violation does not appear. Here we will work in probe limit by neglecting all the reaction of matter fields on the background. The full back reaction will be studied in the forthcoming work~\cite{Caizhang}.  For convenience,  we rescale the value of ATF and its mass and  rewrite $L_2$  to the form
\begin{equation}\label{reL2}
    L_2=-\lambda^2\left[\frac{(dM)^2}{12}+\frac{m^2}4M_{\mu\nu}M^{\mu\nu}+\frac{M^{\mu\nu}F_{\mu\nu}}2+V(M)\right].
\end{equation}
In that case, as one can see, the only difference between the new model and the original one  is that the covariant derivative appearing in \cite{Cai:2014oca} is replaced by the exterior derivative here.
Now we take the AdS-Schwarzschild black brane  as the background geometry with metric,
\begin{equation}\label{metric1}
ds^2=-r^2f(r)dt^2+\frac{dr^2}{r^2f(r)}+r^2(dx^2+dy^2).
\end{equation}
Here $f(r)=1-r_0^3/r^3$ and $r_0$ is the horizon radius. For simplicity, we can set $r_0=1$ in the numerical computation.  Following Ref.~\cite{Cai:2014oca}, we take the following ansatz for matter fields,
\begin{equation}\label{anstz1}
A_\mu=\phi(r)dt+Bxdy,~~M_{\mu\nu}=-p(r)dt\wedge dr+\rho(r)dx\wedge dy.
\end{equation}
In the AdS-Schwarzschild black brane background,  the equations of motion for matter fields read
\begin{equation}\label{totaleq4}
\begin{split}
\rho''+\frac{f'}{f}\rho'-\frac{m^2+V'_\rho}{r^2f}\rho+\frac{B}{r^2f}&=0,\\
(m^2-\frac{V'_p}{r^4})p-\phi'&=0,\\
\phi''+\frac2r\phi'-\lambda^2\left(\frac{p'}4+\frac{p}{2r}\right)&=0,
\end{split}
\end{equation}
where
\begin{equation}\label{VpVrho}
V'_p=\frac{\partial V(M)}{p\partial p},~~V'_\rho=\frac{\partial V(M)}{\rho\partial\rho}.
\end{equation}
As we expect,  $p(r)$ is not a dynamical field and can be directly solved from the second equation in \eqref{totaleq4}.
 With the equations for $p(r)$ and $\phi(r)$, we can obtain
\begin{equation}\label{phip}
\begin{split}
p(r)&=\frac{(1-\lambda^2/4m^2)\sigma}{m^2(1-\lambda^2/4m^2)-V'_p/r^4},\\
\phi(r)&=\sigma(1-\frac{\lambda^2}{4m^2})(1-\frac1r)+\frac{\lambda^2}4\int_1^rp(r) dr.
\end{split}
\end{equation}
Here $\sigma$ is an integration constant, which corresponds to the charge density of the dual system. Assume  that both  $V'_p/r^4$ and $V'_\rho$  decay to zero when $r\rightarrow\infty$, then near the  AdS boundary, we can get the asymptotic solution for $\rho$ as
\begin{equation}\label{asy0}
\rho=\rho_+r^{(1+\delta)/2}+\rho_-r^{(1-\delta)/2}+\cdots+\frac{B}{m^2},
\end{equation}
where $\delta=\sqrt{1+4m^2}$.
%Here we need parameters satisfy the restriction $\lambda^2<4m^2$ so the $\rho$ can condensate when temperature is lower than a critical value~\cite{Caizhang}.

As we discussed  in~\cite{Cai:2014oca}, $\rho(r)$ corresponds to the magnetic response of the dual system. We can define magnetic moment density as,
\begin{equation}\label{defN1}
N=-\lambda^2\int_{r_0}^\infty\frac{\rho}{2r^2}dr.
\end{equation}
In order to have the spontaneous magnetization,   we need to impose the restrictions~\footnote{More details about this will be given in Ref.~\cite{Caizhang}}
\begin{equation}\label{res1}
m^2>\lambda^2/4>0, ~\rho_+=0.
\end{equation}
One can see that in that case, the integration in (\ref{defN1}) converges. If $B\ne0$,  $\rho$ is always nonzero.  So the magnetic moment defined in~\eqref{defN1} is also nonzero, which corresponds to an induced magnetic moment under the external magnetic field. When $B=0$, the leading term is given by $\rho_+$.  According to the AdS/CFT correspondence, $\rho_+$ gives the external source
of the dual operator while $\rho_-$ is its vacuum expectation value.
%In the case of $B=0$, with the regular condition at horizon, the nontrivial solution of $\rho\neq0$ then corresponds to spontaneous magnetization of the dual system.

\begin{figure}
\begin{center}
\includegraphics[width=0.3\textwidth]{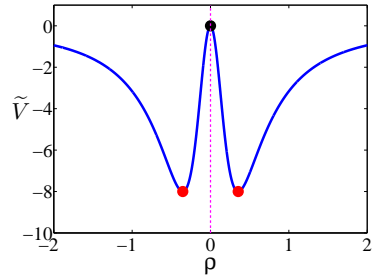}
\caption{A plot of the potential at the horizon $r=r_0=1$. Here $\widetilde{V}=V/[2|J|(1-\lambda^2/4m^2)^2\sigma^2]$ and  the parameters are taken as $m^2=-J=1/8$ and $\lambda=1/2$. The two minimums are at $\rho_c=\pm\sqrt{-m^2/[4J(1-\lambda^2/4m^2)]}=\pm\sqrt{2}/4$, respectively.}
\label{poten1}
\end{center}
\end{figure}

In the case with $V(M)=0$, because the equation for $\rho$ in~\eqref{totaleq4} is linear, there is  no possibility to get a nontrivial condensed solution for $\rho$ when $B=0$. To have a spontaneously condensed solution of $\rho$, we need a self-interaction term for the ATF. One simple form can be taken as
\begin{equation}\label{potential1}
V(M)=\frac{J}8({^*M}_{\mu\nu}M^{\mu\nu})^2.
\end{equation}
Use the ansatz \eqref{anstz1},  we get $V=2Jp^2\rho^2$ so that $V'_p=4J\rho^2$ and $V'_\rho=4Jp^2$. The self-interaction potential in fact is not unique. Here the choice is due to  two reasons. One is that it leads to a simple equation for $p$, through which we can directly express $p$ as a function of $\rho$ and $r$. The other  is that the potential has a  maximum at $\rho=0$ and a minimum at some nonzero positive  $\rho_c$ for fixed $r$. See Fig.~\ref{poten1}.

\begin{figure}
\begin{center}
\includegraphics[width=0.3\textwidth]{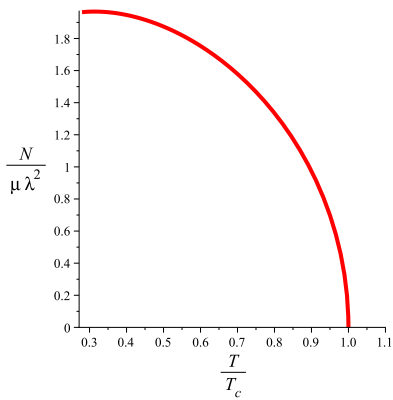}
\caption{The magnetic moment density $N$ as a function of temperature. Here we take the  parameters as $m^2=-J=1/8$ and $\lambda=1/2$. The critical temperature $T_c/\mu\simeq1.7871$}
\label{TNB1}
\end{center}
\end{figure}

In Fig.~\ref{TNB1}, we plot the magnetic moment  density $N$ as a function of temperature. As an typical example, we here take the parameters as $m^2=-J=1/8$ and $\lambda=1/2$. In that case, the critical temperature $T_c/\mu\simeq1.7871$. Thus we see that when temperature is lower than $T_c$, the non-trivial solution of $\rho\neq0$ and spontaneous magnetization appear indeed.

Thus, replacing the  covariant derivative in the kinetic term for the ATF in the original model by the exterior derivative of the ATF, we have shown that the modified model does not include any ghost and
causality violation does not appear.  In the new model,  the spontaneous magnetization and the ferromagnetism/paramagnetism phase transition happen for the ATF in an AdS-Schwarzschild black brane background when
the temperature is lower than its critical temperature.  Furthermore we expect that  all the results presented in~\cite{Cai:2014oca,Cai:2014jta,Cai:2014dza} can be recovered qualitatively with the modified
model.

Next let us stress  why we need the  potential $V(M)$ for the ATF so that the condensation of $\rho$ can happen. As one knows, in some holographic models describing spontaneous condensation such as holographic s-wave superconductor, the condensation can happen without an self-interaction potential.   In the case of the holographic s-wave superconductor model, the complex scalar field couples with
the gauge field, the associated gauge potential can decrease the effective mass squared of the complex scalar field near the horizon but has no any effect on the effective mass squared  far away from the horizon~\cite{Gubser:2008px}, which leads that  the mass squared of the complex scalar can be negative enough  near the horizon so that an instability happens and a nontrivial hair appears.
In   model \eqref{act1},  however, the ATF is real and couples to the gauge field strength,  if $V(M)=0$ and $B=0$,  the instability cannot appear and the condensation will not happen. When we add the term $V(M)$ such as \eqref{potential1},  we can see from the equation of motion of $\rho$ that  the effective mass squared of $\rho$  can be lowed near the horizon, which leads to the expected instability and a nontrivial solution for $\rho$.

In the case of $L_{RM}=0$, we see that in order to have the  instability and the nontrivial solution of $\rho$, we need a self-interaction potential of  ATF.  Such a term could  come from the loop correction of
the ATF in  String/M theory.  In fact, there is an alternative way to generate  the instability, where we can set $V(M)=0$ but $L_{RM}\neq0$. % In this case, ATF has a non-minimum coupling with gravity. This kind of coupling can originate from the quantum correction of gravity.
For example, for simplicity, we can set $a_1\neq0, a_2=a_3=a_4=0$.  In that case,  we can see that the effective mass squared at the horizon and the boundary are,
\begin{equation}\label{meffh}
m^2_{\text{eff}}|_h=m^2+4a_1R_h,~~~m^2_{\text{eff}}|_b=m^2+48a_1,
\end{equation}
respectively. Here the  subscript  $h$ denotes taking the value at  the horizon, while  $b$ for the  value at the AdS boundary. By changing  the scalar curvature near the horizon, we can reach that  the effective mass squared of $\rho$ violates  the Breitenlohner-Freedman bound at the horizon,  but not  at the boundary. This gives us the possibility to generate the instability and spontaneous condensation. %Of course, because the equation for ATF is linear in probe limit, this scheme can realize spontaneously condensing of $\rho$ in case with full backadation. We will investigate it in the future.

\section{summary and discussion}
In this paper, we have started  with  two Casimir invariants of Poinca\'{e} group and  builded  a ghost free and causal ATF  theory step by step. % We first show that the naive thought that an ATF field gives spin-2 particle is wrong.
We have shown that the ATF theory describes spin-1 particles, rather than spin 2 particles.  The naive Lagrangian for ATF  proposed in  Ref.~\cite{Cai:2014oca} includes a vector ghost. To remedy  this,  we have presented a modified  Lagrangian density by replacing the covariant derivative in the kinetic term of the AFT with the exterior derivative.  It turns outs that  the modified Lagrangian density describes  a massive 2-form field. We have argued that  this modified Lagrangian density could be obtained by dimensional compactification of low energy effective action in String/M-theory. For a  general interaction form between ATF and gravity, we have proved that the modified model dose not contain ghost  and does not violate causality in static curved space-time. %Even by taking the non-minimal coupling between the ATF and gravity into account, we show that the modified model is still well.
In AdS-Schwarzschild black brane background, we have  also shown that the spontaneous condensation and paramagnetism/ferromagnetism phase transition can happen in the modified model, and
the main results in Ref.~\cite{Cai:2014oca} keep valid qualitatively.

Note that the main goal of this paper is to present a ghost free and causally well-defined  theory for an ATF.  We believe  that  the self-consistent ATF theory provides a solid base on which various magnetic properties of strongly coupled materials can be investigated in the AdS/CFT framework. We hope to report more significant progress on this model in  the coming works.

%As the background and the interaction with other field are very general, this paper offers a wide foundation about the application of ATF in holographic duality. We hope that more results based on this f%rame could be reported in the future.

It is worth mentioning here that in this paper we neglected these non minimal coupling terms (\ref{LRM1}) between the tensor field and background geometry.  It is natural to ask what would be
the effects and implications  for those terms from the dual boundary theory.  From the equation of motion of the tensor field, we see that those terms provide an effective mass term for the tensor field such that the dimension of the dual tensor operator gets changed if those terms are included. Indeed, as we discussed in the previous section, once those terms are taken into account, even in the case without the potential term of the tensor field, the spontaneous magnetization can happen in the proper choice of the coefficients of those terms.

In addition, let us stress that  our tensor field model looks equivalent to a massive vector field model.  In the mathematical level, it is true when both  the self-interaction of the tensor field and the non-minimal coupling between the tensor field and background geometry vanish. In a general case, they are not equivalent to each other.  From the physical point of view, the magnetic moment is a space-space component of a tensor field and it is a pseudo-vector,  it is therefore more suitable to take a tensor field rather than a vector field in the holographic setup. Another advantage to use the tensor field is that
the time-space component of the tensor field can be viewed as the electric polarization vector and can act as the order parameter for  the paraelectric/ferroelectric phase transition in dielectric materials~\cite{Cai:2014oca}, and the tensor field model can be regarded as a unified holographic model for the paramagnetic/ferromagnetic phase transition and the paraelectric/ferroelectric phase transition.
The other reasons to consider the tensor model  will be presented in \cite{Caizhang}.

{\bf Note added}: After finishing this work, we are informed by Quentin Bailey of the reference \cite{Altschul:2009ae},where a general Lagrangian of an antisymmetric 2-tensor is constructed, even including
non minimal gravitational couplings.  Our action  (\ref{La4} ) just corresponds to the so-called minimal model discussed there.

\section*{Acknowledgements}
This work is finished during one of the authors (R.G. C) visits Kunsan National University, Korea, the warm hospitality extended to him is greatly appreciated.  This work was supported in part by the National Natural Science Foundation of China ( No.11375247 and No.11435006 ).

\appendix.
\section{Computing  $C_2$ for ATF}
\label{app1}
In this appendix let us  compute $C_2$ for a massive ATF. Because  its mass is non vanishing, we can choose a frame where the particle is static and only the spatial components of the spin angular momentum have contribution to $C_2$.  In that case,  we have
\begin{equation}\label{WW2}
C_2=m^2S^iS_i
\end{equation}
where $S^i=\frac12\epsilon^{ijk}S_{jk}$ is the spin vector.  We can get the spin of the massive field by investigating its property under SO(3) transformation. Note that  $S^i$ is an operator of Hilbert space formed by all  ATFs, so does  $C_2$. To compute the eigenvalue of $C_2$, we can choose a suitable basis  of this Hilbert space so that the computation can be done as easy as possible. Of course, the final result is independent of the choice. For convenience, we introduce a new six-components vector $\Psi_A$ ($A=1,2,\cdots,6$) to rearrange the components of ATF such as,
\begin{equation}\label{sixvecr1}
\Psi_A=\left(
\begin{split}
\overrightarrow{P}\\
\overrightarrow{Q}
\end{split}
\right),
\end{equation}
with
\begin{equation}\label{sixvecr2}
\begin{split}
&P_i=M_{0i},~Q_i=\frac12\epsilon_{ijk}M^{jk},\\
&P^i=M^{0i},Q^i=\frac12\epsilon^{ijk}M_{jk}.
\end{split}
~i=1,2,3.
\end{equation}
By definition, we have $P^i=-P_i$ and $Q^i=Q_i$. Under $SO(3)$ transformation, we can see that both $\overrightarrow{P}$ and $\overrightarrow{Q}$ are associated with SO(3) vector. Thus we can find that the representation of $\Psi_A$ under the SO(3) transformation is  $D(1,0)\oplus D(0,1)$.  With  $\Psi_A$, we can easily find the representation of spin operator, which reads
\begin{equation}\label{spin6}
(S^i)_{AB}=\left(
\begin{array}{cc}
J^i&\mathbf{0}\\
\mathbf{0}&J^i
\end{array}
\right)_{AB}
\end{equation}
with $i=1,2,3$ and $A, B=1,2,\cdots,6$. Here $J^i$ are the generators of 3-dimensional spatial rotation. Thus we can find that $(S^iS_i)_{AB}=s(s+1)\delta_{AB}=2\delta_{AB}$, which gives $s=1$, as we expected.

\section{Computing the Hamiltonian for ATF}
\label{app2}
In this appendix, we will show that the Hamiltonian associated with  the Lagrangian~\eqref{La1} is not positive definite  and there is a vector ghost. However, this kind of ghost will not appear in the modified model.

Let us first consider the Lagrangian density \eqref{La1}. From this Lagrangian, we have the canonical momentum density as
\begin{equation}\label{mom1}
\pi^{\mu\nu}=\frac{\partial \mathcal{L}}{\partial\partial_0M_{\mu\nu}}=\partial_0M^{\mu\nu}, ~~\pi_{\mu\nu}=\eta_{\mu\mu'}\eta_{\nu\nu'}\pi^{\mu'\nu'}.
\end{equation}
And the Hamiltonian reads
\begin{equation}\label{Ham0}
H=\int d^3x \mathcal{H}
\end{equation}
with the Hamiltonian density $\mathcal{H}$,
\begin{equation}\label{Ham1}
\begin{split}
&\mathcal{H}=\frac12\pi^{\mu\nu}\partial_0M_{\mu\nu}-\mathcal{L}\\
&=\frac14\left[\pi^{\mu\nu}\pi_{\mu\nu}+(\partial_iM_{\mu\nu})\partial^iM^{\mu\nu} +m^2M^{\mu\nu}M_{\mu\nu}\right].
\end{split}
\end{equation}
Now let us adopt the two vectors $\overrightarrow{P}$ and $\overrightarrow{Q}$ defined in~\eqref{sixvecr2} to rewrite  the Hamiltonian. From the expression \eqref{mom1}, we have their canonical momentum densities
\begin{equation}\label{PN1b}
\begin{split}
&\pi^i(P)=\pi^{0i},~~~\pi^i(Q)={\epsilon^i}_{jk}\pi^{jk},\\
&\pi_i(P)=\pi_{0i},~~~\pi_i(Q)=\epsilon_{ijk}\pi^{jk}.
\end{split}
\end{equation}
We see that $\pi^i(P)=-\pi_i(P)$ and $\pi^i(Q)=\pi_i(Q)$. Note that here we have just used some new notations to represent the components of $M_{\mu\nu}$ and $\pi_{\mu\nu}$, which does  not change any physics of the theory. The Hamiltonian then decouples into two parts associated with two vector fields under SO(3) transformation,
\begin{equation}\label{PN3}
H=H_P+H_Q,
\end{equation}
where
\begin{equation}\label{HP1}
\begin{split}
H_P=-\frac12\sum_{i=1}^3\int d^3x[\pi^i(P)\pi^i(P)+(\overrightarrow{\nabla} P_i)^2 +m^2P_i^2],\\
H_Q=\frac12\sum_{i=1}^3\int d^3x[\pi^i(Q)\pi^i(Q)+(\overrightarrow{\nabla} Q_i)^2 +m^2Q_i^2].
\end{split}
\end{equation}
We see that the part for $P$ is negative definite. More precisely speaking, the Hamiltonian of the theory has no lower bound, which means that there exist  infinite ghost states, but no ground state  in Fock's space.

On the other hand, if we take  the Lagrangian density such as \eqref{La3} or \eqref{La4}, the result is rather different. To make it clear, we adopt  the  Lagrangian density \eqref{La4}. Taking the constraint Eq.~\eqref{tran1} into account, we find the Hamiltonian can be written as the same form as \eqref{HP1}.  However, the canonical momentum density for $P$ now is zero, i.e. $\pi^i(P)=0$, and the components $P_i$ are not independent variables.  With the equation \eqref{eom2b}, we see that the equation of motion of $P$ reads
\begin{equation}\label{eomp1}
0=\partial^0\pi^i(P)+\overrightarrow{\nabla}^2P_i-m^2P_i=\overrightarrow{\nabla}^2P_i-m^2P_i.
\end{equation}
Put it into the expression for $H_P$, we have
\begin{equation}\label{HP2}
\begin{split}
H_P&=-\frac12\sum_{i=1}^3\int d^3x[(\overrightarrow{\nabla} P_i)^2 +P_i\overrightarrow{\nabla}^2P_i]\\
&=-\frac12\sum_{i=1}^3\int d^3x\overrightarrow{\nabla}\cdot(P_i\overrightarrow{\nabla} P_i),
\end{split}
\end{equation}
which is just a total divergence term and can be removed from the Hamiltonian of the system. As a result we get $H=H_Q$, which is obviously positive definite. We see that there are 3 polarization states for every momentum. In addition, if quantize this system,  we can see that the projection of spin along the direction of momentum can be $\pm1, 0$. All  three polarizations are physical.

\section{Giving the $p$-form field mass by compactification}
\label{app3}
As we have argued  in  section \ref{sec1}, the modified ATF theory  has a clear physical origin, which can be treated as the low energy limit of some fundamental field, such as $p$-form field,  in String/M theory.  In String/M theory, however, the $p$-form field is massless. Therefore  we need to clarify there is a suitable mechanism to give the mass term. In fact, there are a few  mechanisms for this as we mentioned in section \ref{sec1}. Here we give a very simple mechanism to give the mass term by the generalized  KK dimensional reduction.
%which is just based on the fact the AdS/CFT duality comes from the the compactification of $AdS_5\times S^5$ in the type II-B superString theory.
%We will discuss a simplified model, in which
%For simplicity, here we assume that  there is only one compacted dimension.  It is trivial to generalize to the case with more compacted dimensions.
This kind of dimensional reduction  is discussed in Ref.~\cite{Fu:2012sa}.  Here we just give a brief introduction.

For simplicity, we set $\mu_p=0$ in  action \eqref{RRsect} and take the background metric as
\begin{equation}\label{cmetric1}
ds^2=e^{2A(z)}(h_{\mu\nu}dx^\mu dx^\nu+dz^2),
\end{equation}
where $z$ denotes the extra dimension and $A$ is a function of $z$. Then the equations of motion of the action \eqref{RRsect} in the conformal metric \eqref{cmetric1} read
\begin{equation}\label{eomc1}
\begin{split}
\partial_{\mu_1}(\sqrt{-g}E^{\mu_1\mu_2\cdots\mu_{p+1}})+\partial_z(\sqrt{-g}E^{z\mu_2\cdots\mu_{p+1}})&=0,\\
\partial_{\mu_1}(\sqrt{-g}E^{\mu_1\mu_2\cdots\mu_{p}z})&=0,
\end{split}
\end{equation}
where $E_{\mu_1\mu_2\cdots\mu_{p+1}}=(dC)_{\mu_1\mu_2\cdots\mu_{p+1}}$. Consider  the gauge symmetry, we can set $C_{\mu_1\mu_2\cdots\mu_{p-1}z}=0$. Next we make a decomposition for $C_{\mu_1\mu_2\cdots\mu_p}$ as
\begin{equation}\label{decomp1}
C_{\mu_1\mu_2\cdots\mu_p}=\sum_n\widetilde{C}^{(n)}_{\mu_1\mu_2\cdots\mu_p}K_n(z)e^{A(z)/2}.
\end{equation}
By this decomposition, the field strength $E_{\mu_1\mu_2\cdots\mu_{p+1}}$ can be expressed as
\begin{equation}\label{decomp2a}
E_{\mu_1\mu_2\cdots\mu_{p+1}}(x^\mu,z)=\sum_n\widetilde{E}^{(n)}_{\mu_1\mu_2\cdots\mu_{p+1}}K_n(z)e^{A(z)/2}
\end{equation}
and
\begin{equation}\label{decomp2b}
\begin{split}
&E_{\mu_1\mu_2\cdots\mu_{p}z}(x^\mu,z)\\
&=\sum_n\widetilde{E}^{(n)}_{\mu_1\mu_2\cdots\mu_{p+1}}(K'_n+K_nA'_n/2)(z)e^{A(z)/2}
\end{split}
\end{equation}
with $\widetilde{E}^{(n)}_{\mu_1\mu_2\cdots\mu_{p+1}}=\partial_{[\mu_1}\widetilde{C}^{(n)}_{\mu_2\mu_3\cdots\mu_{p}]}$. Substituting  these decompositions into the equations of motion~\eqref{eomc1}, we obtain the equation for the function $K_n(z)$, which has a Schr\"{o}dinger-like form  as,
\begin{equation}\label{schr1}
\left[-\partial^2_z+\frac{{A'}^2}4+\frac{A''}2\right]K_n=m_n^2K_n.
\end{equation}
The equation for $K_n$ is second order ODE. The properties of its solutions depend on the function $A(z)$.  We further assume that the solutions of \eqref{schr1} with some suitable boundary conditions form a complete orthogonal basis with the function basis ${K_n(z)}$. By a straightforward calculation, we can show that  the the effective action of the $p$-form field  reads,
\begin{equation}\label{effC1}
\begin{split}
S_{\text{eff}}&=\sum_n\int d^{D-1}x\sqrt{-h}\left[\widetilde{E}^{(n)\mu_1\mu_2\cdots\mu_{p+1}}\widetilde{E}^{(n)}_{\mu_1\mu_2\cdots\mu_{p+1}}\right.\\ &+\left.\frac{m_n^2}{p+1}\widetilde{C}^{(n)\mu_1\mu_2\cdots\mu_p}\widetilde{C}^{(n)}_{\mu_1\mu_2\cdots\mu_p}\right].
\end{split}
\end{equation}
We see that for the excited states of KK modes, we have $m_n^2\neq0$, which gives the mass for $p$-form in a lower dimension. By specializing to the case $p=2$, it gives a mass term for a massless 2-form field.  Here we further mention that if $A=0$ in (\ref{cmetric1}),  the dimension along $z$ is required to be compact. In this case, the above dimensional reduction is just the standard KK one. On the
other hand, if $A\ne 0$, the dimension along $z$ is not required to be compact.  This is called warped dimensional reduction (compactification).

%%%%%%%%%%%%%%%%%%%%%%%%%%%%%%%%

%%%%%%%%%%%%%%%%%%%%

%%%%%%%%%%%%%%%%%%%%%%

\begin{thebibliography}{99}
%\cite{Maldacena:1997re}
\bibitem{Maldacena:1997re}
  J.~M.~Maldacena,
  ``The large N limit of superconformal field theories and supergravity,''
  Int.\ J.\ Theor.\ Phys.\  {\bf 38}, 1113 (1999)
  %Adv.\ Theor.\ Math.\ Phys.\  {\bf 2}, 231 (1998)
  [arXiv:hep-th/9711200].
  %%CITATION = IJTPB,38,1113;%%

%\cite{Gubser:1998bc}
\bibitem{Gubser:1998bc}
  S.~S.~Gubser, I.~R.~Klebanov and A.~M.~Polyakov,
  ``Gauge theory correlators from non-critical string theory,''
  Phys.\ Lett.\  B {\bf 428}, 105 (1998) [arXiv:hep-th/9802109].
  %%CITATION = PHLTA,B428,105;%%



%\cite{Witten:1998qj}
\bibitem{Witten:1998qj}
  E.~Witten,
  ``Anti-de Sitter space and holography,''
  Adv.\ Theor.\ Math.\ Phys.\  {\bf 2}, 253 (1998)
  %[arXiv:hep-th/9802150].
  %%CITATION = 00203,2,253;%%

%\cite{Witten:1998qj2}
\bibitem{Witten:1998qj2}
E. Witten,
``Anti-de Sitter space, thermal phase transition, and confinement in gauge theories",
Adv. Theor. Math. Phys. {\bf 2}, 505 (1998)
[arXiv:hep-th/9803131].

%\cite{Hartnoll:2008vx}
\bibitem{Hartnoll:2008vx}
  S.~A.~Hartnoll, C.~P.~Herzog and G.~T.~Horowitz,
  ``Building a Holographic Superconductor,''
  Phys.\ Rev.\ Lett.\  {\bf 101}, 031601 (2008)
  [arXiv:0803.3295 [hep-th]].
  %%CITATION = ARXIV:0803.3295;%%
  %527 citations counted in INSPIRE as of 05 Jun 2013

%\cite{Lee:2008xf}
\bibitem{Lee:2008xf}
  S.~S.~Lee,
  ``A Non-Fermi Liquid from a Charged Black Hole: A Critical Fermi Ball,''
  Phys.\ Rev.\ D {\bf 79}, 086006 (2009)
  [arXiv:0809.3402 [hep-th]].
  %%CITATION = ARXIV:0809.3402;%%
  %214 citations counted in INSPIRE as of 11 Dec 2014

%\cite{Liu:2009dm}
\bibitem{Liu:2009dm}
  H.~Liu, J.~McGreevy and D.~Vegh,
  ``Non-Fermi liquids from holography,''
  Phys.\ Rev.\ D {\bf 83}, 065029 (2011)
  [arXiv:0903.2477 [hep-th]]
  %%CITATION = ARXIV:0903.2477;%%
  %237 citations counted in INSPIRE as of 05 Jun 2013

 %\cite{Cubrovic:2009ye}
\bibitem{Cubrovic:2009ye}
  M.~Cubrovic, J.~Zaanen and K.~Schalm,
  ``String Theory, Quantum Phase Transitions and the Emergent Fermi-Liquid,''
  Science {\bf 325}, 439 (2009)
  [arXiv:0904.1993 [hep-th]].
  %%CITATION = ARXIV:0904.1993;%%
  %253 citations counted in INSPIRE as of 11 Dec 2014

%\cite{Donos:2013gda}
\bibitem{Donos:2013gda}
  A.~Donos and J.~P.~Gauntlett,
  ``Holographic charge density waves,''
  Phys.\ Rev.\ D {\bf 87}, 126008 (2013)
  [arXiv:1303.4398 [hep-th]].
  %%CITATION = ARXIV:1303.4398;%%
  %24 citations counted in INSPIRE as of 16 mar 2015

%\cite{Ling:2014saa}
\bibitem{Ling:2014saa}
  Y.~Ling, C.~Niu, J.~Wu, Z.~Xian and H.~b.~Zhang,
  ``Metal-insulator Transition by Holographic Charge Density Waves,''
  Phys.\ Rev.\ Lett.\  {\bf 113}, 091602 (2014)
  [arXiv:1404.0777 [hep-th]].
  %%CITATION = ARXIV:1404.0777;%%
  %11 citations counted in INSPIRE as of 16 Mar 2015

%\cite{Cai:2015cya}
\bibitem{Cai:2015cya}
  R.~G.~Cai, L.~Li, L.~F.~Li and R.~Q.~Yang,
  ``Introduction to Holographic Superconductor Models,''
  Sci.\ China Phys.\ Mech.\ Astron.\  {\bf 58}, 060401 (2015)
  [arXiv:1502.00437 [hep-th]].
  %%CITATION = ARXIV:1502.00437;%%
  %3 citations counted in INSPIRE as of 31 Mar 2015



%\cite{Amoretti:2013oia}
\bibitem{Montull:2009}
 M.~Montull, A.~Pomarol and P.~J.~Silva,
  ``The Holographic Superconductor Vortex,''
  Phys.\ Rev.\ Lett.\  {\bf 103}, 091601 (2009)
  [arXiv:0906.2396 [hep-th]].
  %%CITATION = ARXIV:0906.2396;%%
  %63 citations counted in INSPIRE as of 16 Oct 2014

%\cite{Donos:2012yu}
\bibitem{Donos:2012yu}
  A.~Donos, J.~P.~Gauntlett, J.~Sonner and B.~Withers,
  ``Competing orders in M-theory: superfluids, stripes and metamagnetism,''
   JHEP {\bf 1303}, 108 (2013)  [arXiv:1212.0871 [hep-th]].  %%CITATION = ARXIV:1212.0871;%%  %22 citations counted in INSPIRE as of 29 Mar 2014



%\cite{Albash:2008eh}
\bibitem{Albash:2008eh}
  T.~Albash and C.~V.~Johnson,
  ``A Holographic Superconductor in an External Magnetic Field,''
   JHEP {\bf 0809}, 121 (2008)  [arXiv:0804.3466 [hep-th]].  %%CITATION = ARXIV:0804.3466;%%  %109 citations counted in INSPIRE as of 28 Mar 2014


%\cite{Montull:2012fy}
\bibitem{Montull:2012fy}
  M.~Montull, O.~Pujolas, A.~Salvio and P.~J.~Silva,
``Magnetic Response in the Holographic Insulator/Superconductor Transition,''
 JHEP {\bf 1204} (2012) 135
 [arXiv:1202.0006 [hep-th]].  %%CITATION = ARXIV:1202.0006;%%  %27 citations counted in INSPIRE as of 22 Mar 2014



%\cite{Iqbal:2010eh}
\bibitem{Iqbal:2010eh}
  N.~Iqbal, H.~Liu, M.~Mezei and Q.~Si,
  ``Quantum phase transitions in holographic models of magnetism and superconductors,''
    Phys.\ Rev.\ D {\bf 82}, 045002 (2010)
    [arXiv:1003.0010 [hep-th]].  %%CITATION = ARXIV:1003.0010;%%  %64 citations counted in INSPIRE as of 22 Mar 2014



%\cite{Kondo}
\bibitem{Kondo}
J. Kondo,  ``Resistance Minimum in Dilute Magnetic Alloys",
 Prog. Theor. Phys. {\bf 32}, 37 (1964)


%\cite{A.P.Ramirez:1995ge}
\bibitem{A.P.Ramirez:1995ge}
  A.~P.~Ramirez,
  ``Colossal magnetoresistance,''
  J. Phys.: Condens. Matter {\bf 9} (1997) 8171-8199.

%\cite{Aoki}
\bibitem{Aoki}
D. Aoki, A. Huxley, E. Ressouche, D. Braithwaite, J. Flouquet, J-P. Brison, E. Lhotel and C. Paulsen, ``Coexistence of superconductivity and ferromagnetism in URhGe", Nature {\bf 413} 613 (2001):

%\cite{Huy}
\bibitem{Huy}
N. T. Huy, A. Gasparini, et al., ``Superconductivity on the border of weak itinerant ferromagnetism in UCoGe", Phys. Rev. Lett.  {\bf 99}, 067006 (2007)

%\cite{Cai:2014oca}
\bibitem{Cai:2014oca}
  R.~G.~Cai and R.~Q.~Yang,
  ``Paramagnetism-Ferromagnetism Phase Transition in a Dyonic Black Hole,''
  Phys.\ Rev.\ D {\bf 90}, no. 8, 081901 (2014)
  [arXiv:1404.2856 [hep-th]].
  %%CITATION = ARXIV:1404.2856;%%
  %10 citations counted in INSPIRE as of 15 Mar 2015

%\cite{Cai:2014jta}
\bibitem{Cai:2014jta}
  R.~G.~Cai and R.~Q.~Yang,
  ``A Holographic Model for Paramagnetism/antiferromagnetism Phase Transition,''
  [arXiv:1404.7737 [hep-th]].
  %%CITATION = ARXIV:1404.7737;%%
  %5 citations counted in INSPIRE as of 15 Mar 2015

%\cite{Cai:2014dza}
\bibitem{Cai:2014dza}
  R.~G.~Cai and R.~Q.~Yang,
  ``Coexistence and competition of ferromagnetism and $p$-wave superconductivity in holographic model,''
  Phys.\ Rev.\ D {\bf 91}, no. 2, 026001 (2015)
  [arXiv:1410.5080 [hep-th]].
  %%CITATION = ARXIV:1410.5080;%%
  %3 citations counted in INSPIRE as of 15 mar 2015

%\cite{Deser:2000dz}
\bibitem{Deser:2000dz}
  S.~Deser, A.~Waldron, and V.~Pascalutsa, 
  ``Massive spin 3/2 electrodynamics,''  Phys.\ Rev.\ D {\bf 62}, 105031 (2000)  [hep-th/0003011].  %%CITATION = HEP-TH/0003011;%%  %62 citations counted in INSPIRE as of 16 Mar 2015


%\cite{Buchbinder:2000fy}
\bibitem{Buchbinder:2000fy}
  I.~L.~Buchbinder, D.~M.~Gitman and V.~D.~Pershin,
  ``Causality of massive spin-2 field in external gravity,''  Phys.\ Lett.\ B {\bf 492}, 161 (2000)  [hep-th/0006144].  %%CITATION = HEP-TH/0006144;%%  %66 citations counted in INSPIRE as of 14 mar 2015

%\cite{Rahman:2013sta}
\bibitem{Rahman:2013sta}
  R.~Rahman,
  ``Higher Spin Theory - Part I,''  PoS ModaveVIII 004 (2012)  [arXiv:1307.3199 [hep-th]].  %%CITATION = ARXIV:1307.3199;%%  %2 citations counted in INSPIRE as of 29 Mar 2015


%%%%%%%%%%%%%%%%%


%\cite{H.Ruegg}
\bibitem{H.Ruegg}
H. Ruegg and M. Ruiz-Altaba, ``The Stueckelberg field," Int. J. Mod. Phys. {\bf A19}
3265-3348 (2004) , [hep-th/0304245].

%\cite{B.Kors}
\bibitem{B.Kors}
B. Kors and P. Nath, ¡°Aspects of the Stueckelberg extension,¡± JHEP {\bf 07} 069 (2005), [hep-ph/0503208].

%\cite{Allen}
\bibitem{Allen}
T. J. Allen, M. J. Bowick, and A. Lahiri, ¡°Topological mass generation in
(3+1)-dimensions,¡±Mod. Phys. Lett. {\bf A6} 559-572 (1991).

%\cite{Fu:2012sa}
\bibitem{Fu:2012sa}
  C.~E.~Fu, Y.~X.~Liu, K.~Yang and S.~W.~Wei,
  ``q-Form fields on p-branes,''
  JHEP {\bf 1210}, 060 (2012)
  [arXiv:1207.3152 [hep-th]].
  %%CITATION = ARXIV:1207.3152;%%
  %9 citations counted in INSPIRE as of 11 mar 2015



%\cite{Caizhang}
\bibitem{Caizhang}
  R.~G.~Cai, R.~Q.~Yang, C.~Y.~Zhang, in preperation.



%\cite{Gubser:2008px}
\bibitem{Gubser:2008px}
  S.~S.~Gubser,
  ``Breaking an Abelian gauge symmetry near a black hole horizon,''
  Phys.\ Rev.\ D {\bf 78}, 065034 (2008)
  [arXiv:0801.2977 [hep-th]].
  %%CITATION = ARXIV:0801.2977;%%
  %536 citations counted in INSPIRE as of 15 Mar 2015

%\cite{Altschul:2009ae}
\bibitem{Altschul:2009ae}
  B.~Altschul, Q.~G.~Bailey and V.~A.~Kostelecky,
  ``Lorentz violation with an antisymmetric tensor,''
  Phys.\ Rev.\ D {\bf 81}, 065028 (2010)
  [arXiv:0912.4852 [gr-qc]].
  %%CITATION = ARXIV:0912.4852;%%
  %48 citations counted in INSPIRE as of 23 Apr 2015

\end{thebibliography}
\end{document}